\documentclass[
    ,final            
  ]
  {aipproc}

\layoutstyle{8x11single}

\def\be{\begin{eqnarray} &&}

\def\ee{\end{eqnarray}}

\def\oneh{{\textstyle {1\over 2}}}
\def\ket#1{\hbox{$\vert #1\rangle$}}   
\def\bra#1{\hbox{$\langle #1\vert$}}   

\begin{document}

\title{Longitudinal and Transverse Parton Momentum Distributions for Hadrons  
within Relativistic Constituent Quark Models
}

\classification{12.39.Ki, 11.10.St, 13.40.-f, 14.40.Aq}
\keywords      {hadron, parton distributions}

\author{T. Frederico}{
  address={Dep. de F\'\i sica, Instituto Tecnol\'ogico de Aeron\'autica,
 12.228-900 S\~ao Jos\'e dos
Campos, S\~ao Paulo, Brazil}
}

\author{E.Pace}{
  address={Dipartimento di Fisica, Universit\`a degli Studi di Roma "Tor Vergata" and Istituto
Nazionale di Fisica Nucleare, Sezione Tor Vergata, Via della Ricerca
Scientifica 1, 00133  Roma, Italy
}
}
\author{B.Pasquini}{
  address={Dipartimento di Fisica Nucleare e Teorica, Universit\`a degli
Studi di Pavia and Istituto Nazionale di Fisica Nucleare, Sezione di
Pavia, Pavia, Italy}
}

\author{G. Salm\`e}{
  address={Istituto  Nazionale di Fisica Nucleare, Sezione di Roma, P.le A. Moro 2,
 00185 Roma, Italy}
}

\begin{abstract}
 Longitudinal and transverse parton distributions for  pion and  nucleon 
 are calculated from hadron vertexes
 obtained by a study of  
form factors within relativistic
  quark models. 
 The relevance of the
 one-gluon-exchange dominance at short range for the behavior of the form factors at large
 momentum transfer and of the parton distributions at the end points is stressed.
\end{abstract}

\maketitle


\section{INTRODUCTION}
A detailed description of  hadron structure can be accessed through the
generalized parton
distributions (GPD's)
(see, e.g., Refs. \cite{Ji98,diehlpr,pasq}). 
We determine  hadron vertex functions 
by a study of 
electromagnetic (em) form factors (ff) in the spacelike (SL) region within
constituent  quark models, and then we use the  vertex
functions to evaluate the GPD's. 

In this contribution we shortly review :
(i) our results for the unpolarized GPD's
 of the pion within covariant and light-front (LF) models \cite{FPPS};
and (ii) our preliminary results for the unpolarized longitudinal
and transverse parton momentum distributions (TMD) in the nucleon within
a LF framework \cite{PDFPS}.

In order to study the GPD's in the valence region as well as in the nonvalence (NV) region,
the Fock state decomposition \cite{Bro} of the hadron state  
has to be considered : e.g. for the pion
$ ~| \pi \rangle ~~=  ~~
|q\bar{q} \rangle ~~ + 
~~ |q \bar{q} ~ q \bar{q}\rangle ~ + ~
|q \bar{q} ~g\rangle ~ + ~... ~$.

Isoscalar (IS) and isovector (IV) pion GPD's, $H^{0,1}_{\pi}(x,\xi,t)$, in
the light-cone gauge are
\be
\hspace{-0.7 cm}  H^{0}_{\pi} =  \int \frac{dz^-}{4\pi} e^{i x P^+ z^-}
\left . \langle p' | ~\bar \psi_q
(-\frac{z}{2} ) ~ \gamma ^+   \psi_q (\frac{z}{2} ) | p
\rangle \right |_{\tilde z=0} 
\hspace{0.6 cm}  H^{1}_{\pi} = \hspace{-0.1 cm}\int \frac{dz^-}{4\pi} e^{i x P^+ z^-}
 \left . \langle p' | \bar \psi_q
(-\frac{z}{2})  \gamma ^+  \tau_3  \psi_q (\frac{z}{2}) 
  | p \rangle
\right |_{\tilde z=0}
\ee
where $\tilde z \equiv \{z^+ = z^0 + z^3 , {\bf z}_\perp\}$  , and
  $\psi_q(z)$ is the quark  field  isodoublet, while
  \begin{figure}[b]
\vspace{-0.7cm}
\hspace{.3cm}
{\includegraphics[width=6.cm]{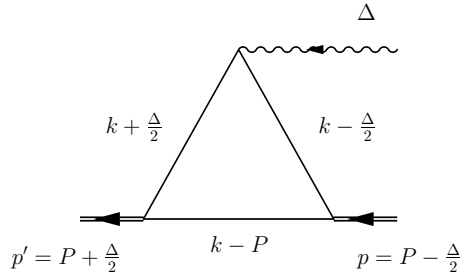}}
\caption{Diagrammatic representation of the pion GPD, with four momentum definitions 
(after Ref. \cite{FPPS}).}
\end{figure}
  $ ~{{P}}=\frac12(p'+p)$,
  $ ~\Delta = p^\prime-p$,
 $~x=k^+/{P}^+$ $~\xi=-(p'^+ - p^+)/2 {{P}}^+$,
 $~t=\Delta^2~$, 
with $k$ is the average
momentum of the active quark, i.e. the one that interacts with the photon (see Fig. 1).
The variable $x$  allows one to single out (i) the valence region (DGLAP \cite{dglap}), for
$~1\ge x\ge |\xi|$ and $-|\xi| \ge x\ge -1$, diagonal in the Fock space; 
and (ii) the nonvalence region (ERBL \cite{erbl}), for  $|\xi|> x >-|\xi|$,  
non diagonal in the Fock space.
Three pion models are used :
\begin{description}
\item
  [An analytic covariant pion model with symmetric regulators and a bare photon vertex.]
 
 We use a pion Bethe-Salpeter amplitude  (BSA)  with a $\gamma^5$ coupling \cite{tob92}
and a constituent quark (CQ) mass $m=220$ MeV.
Two  covariant symmetric forms for the momentum dependent part, $\Lambda(k-P,p)$, of the 
BSA are considered:
 i) a sum form,  and ii) a product form,
which depend on a parameter 
chosen to fit the pion   decay 
constant $f_\pi$  in each model.
The $u$-quark GPD in the pion, $~H^u(x,\xi,t) = H^{I=0}_{\pi}$ $+$  $H^{I=1}_{\pi}$~,
 is given in {\em{impulse approximation}} by
\be
\hspace{-.7cm} H^u(x,\xi,t) = -\imath ~N_c~2 ~ \frac {m^2}{f^2_\pi} ~ 
 \int
\frac{d^4k}{2(2\pi)^4} ~ \delta(P^+x-k^+) \; ~ V^+ ~
 \Lambda(k-P,p^{\prime})\; \;  
\Lambda(k-P,p)  \quad ,
\label{jmu}
\ee
where 
the $\delta$ function imposes the correct support,  $-|\xi| \le x \le 1$, 
for the active quark,
 $N_c=3~$ is the number of colors, 
 $S(k)$  the free quark propagator and
 $
 V^+=Tr\left \{ S\left({k}-{P}\right)
\gamma^5 ~S\left({k}+
{\Delta}/{2}\right) 
\gamma^+~S\left({k}-{\Delta}/{2}\right)\gamma^5\right \}
 \label{trace}
$.
We adopt a Breit frame with $~~\Delta^+ = -\Delta^- \ge0$. Then in this model
 the whole kinematical range $-1 \le \xi \le 1$ can be explored. 
\item
 [Mandelstam-inspired pion light-front model.]
 
We extend to GPD's \cite{FPPS} the model \cite{DFPS} for the pion  ff,
based on the covariant
Mandelstam formula for the current \cite{mandel}, with
a microscopic Vector Meson (VM) dominance dressing for the photon vertex. 
The   expression of Eq. (\ref{jmu}) for  $ H^u$ holds, but
the bare quark-photon vertex, $\gamma^+$,
is replaced by the VM dominance (VMD) vertex of \cite{DFPS}.
  In the $k^-$ integration only the  propagators poles  are considered, i.e.  the BSA
 analytic structure  is disregarded in  i)  the  pion state and ii) 
 the  photon vertex.
 The dynamical inputs are the pion and VM LF wave functions (wf's).
 For the tridimensional reduction of  
 the $n-th$ VM BSA in the valence sector, $0<k^+<P^+_n$, we take the 
 eigenfunction of a relativistic  
mass operator \cite{Fred},  normalized to the probability
of the valence Fock state \cite{DFPS}.
 For the pion in the valence sector 
the  
 eigenfunction of the mass operator  \cite{Fred} is used, while
 for the {\em{NV pion vertex}},
 a constant is assumed  \cite{Choi}. 
All of the parameters of \cite{DFPS} are used, but for 
the CQ mass $m= 200$ MeV, instead of $m= 265$ MeV. A parameter, $w=-1$, which
  modulates the relative weights of our two instantaneous contributions is used 
  to fit the pion ff.
  For this model we take the Breit frame where $ {\Delta}_\perp = 0$, 
and assume $m_\pi = 0$ (see Ref. \cite{DFPS}). 
Then $\xi = -1$ and only the NV region
contributes.
 
\item
  [Light-front Hamiltonian dynamics model.]

In the light-front Hamiltonian dynamics (LFHD) model \cite{Chung}
 the Drell-Yan $~\Delta^+ = 0~$ reference frame is adopted
and then the variable
$x$ becomes the longitudinal momentum fraction $x_q$, since $\xi=0$ for any $t$.
Within the LFHD 
 model the pion LF wave function is
\be
 \hspace{-0.8cm}\Psi_\pi  (x,{\bf \kappa}_{\perp }; \lambda_q, \lambda_{\bar q})
= \psi_\pi(x,{\bf \kappa}_{\perp })  ~ 
\sum_{\mu_q,\mu_{\bar q}} \left(\oneh\mu_q\oneh \mu_{\bar q}|00\right)
{D}^{1/2\,*}_{\mu_q\lambda_q}\left[R({\bf \kappa})\right] 
{D}^{1/2\,*}_{\mu_{\bar q}\lambda_{\bar q}}\left[R(-{\bf \kappa})\right] \quad ,
\label{pionwf}
\ee
with $~\lambda_i~$ ($i=q, \bar q$)~  the   spin projections, 
$~{\bf \kappa}~	\equiv ~\{{\bf \kappa}_\perp, ~ \kappa_z\} $,
$\kappa_z = M_0(x,{\bf \kappa}_\perp)~(x -{1 \over 2})$ and
$M_0(x,{\bf \kappa}_\perp)$ the pion free mass.

The  Melosh rotation  
${D}^{1/2}_{\lambda\mu}\left[R({\bf \kappa})\right ] =
\bra{\lambda}R({\bf \kappa})\ket{\mu} 
$
converts the  instant-form spins  
into LF spins and ensure the rotational covariance. 
For the    momentum component of the pion wf we use
a Gaussian form. 
The CQ mass $m=250$ MeV and a parameter 
 in the exponent
are adjusted to fit $f_{\pi}$ and the pion charge radius \cite{FPPS}.
In this model the pion GPD $H^u(x,\xi=0,t) $ in the range $0 \le x \le 1$ is given by
a diagonal contribution with $n=2$ constituents
\be
\hspace{-0.8cm}H^u
 = \sum_{\{\lambda_i\}}
\int
{d{\bf \kappa}_\perp \over 2(2\pi)^3}~
\Psi_\pi^{*}(x,{\bf
\kappa}^\prime_{\perp};\{\lambda_i\})
\Psi_\pi(x,{\bf \kappa}_{\perp};\{\lambda_i\}) ~~.
\label{eq:overlap}
\ee
 Initial and final
transverse components of  active quark momenta
in the intrinsic frame  are related by
${\bf \kappa^\prime}_{\perp } = 
{\bf \kappa}_{\perp } + (1-x) ~ { \Delta}_{\perp}$.
Then at  large $|t|$, i.e. at large  $|\Delta_{\perp}|$, $H^u(x,\xi=0,t) $
is expected to be non vanishing only for $x \sim 1$.
\end{description}
\section{Pion longitudinal and transverse momentum distributions}
\begin{figure}[t]
\vspace{0.5cm}
\includegraphics[width=7.cm]{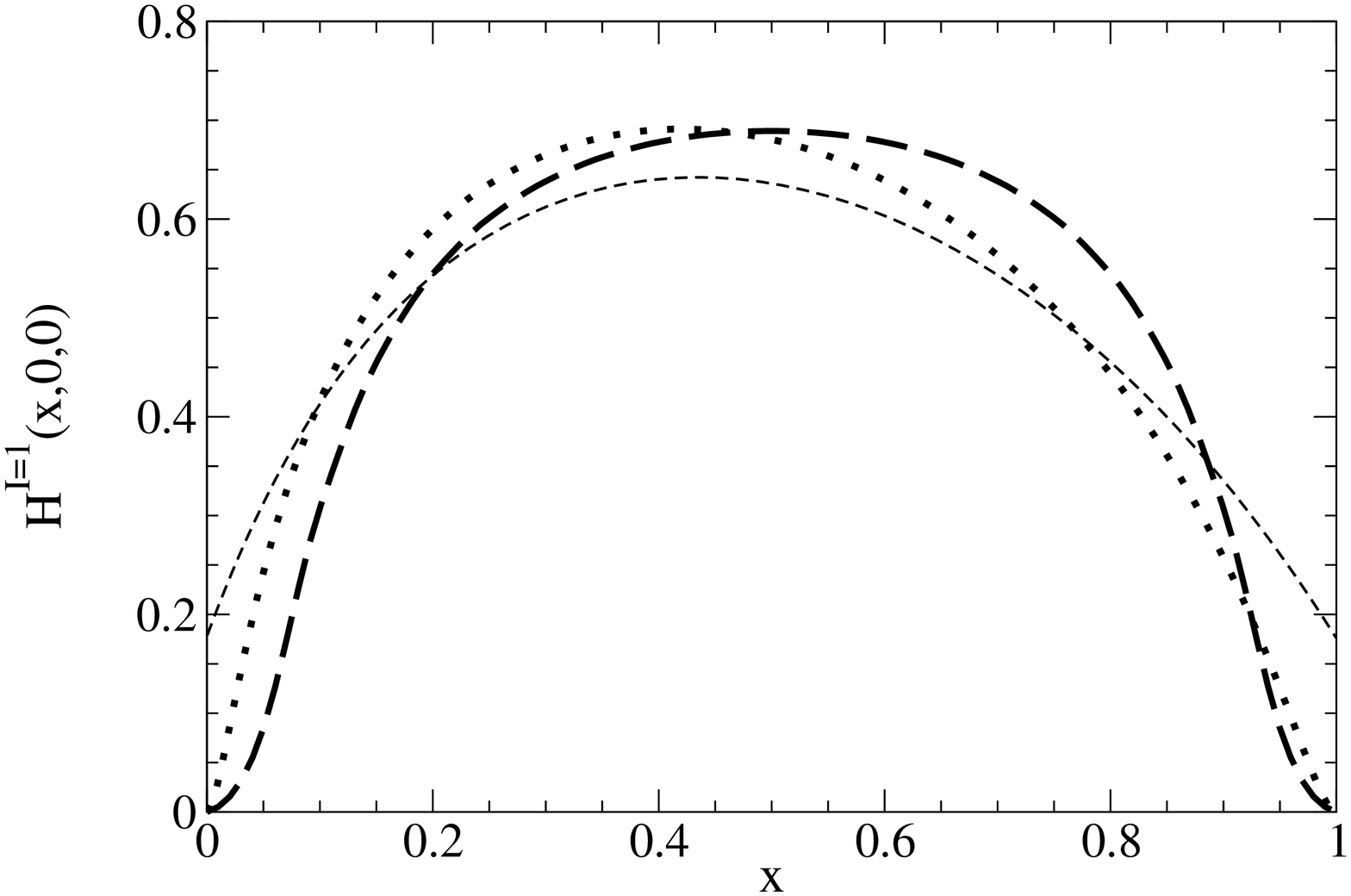}
\caption{ Thin dashed line:
covariant model with the sum-form BSA. 
Dotted line: covariant model with  the product-form BSA.
Thick dashed line: LFHD model 
 (after Ref. \cite{FPPS}).}
\label{strucx} 
\end{figure}
At $~\xi=0~$ one has $~x~= ~ x_q~$ and for $~t=0~$ one gets from $~H^u~$ the longitudinal
 momentum distribution
\be 
\hspace{-0.2cm} u(x) = H^u(x,0,0) =  2 ~ H^{I=1}_\pi(x,0,0)
= \int d{\bf k}_\perp~ f_1(x,k_\perp) \quad \quad (x\ge 0)~ ,
\label{f1xk}
~~\ee
where  
$f_1(x,{k}_\perp)$ is the TMD.
The  $u(x)$ distributions for our models are compared in Fig. 2.
The covariant sum-form model is unable to give a vanishing value 
 for $u(x)$ at the end points, while
 the product-form model  describes
both  the ff tail  and  the end-point fall-off of the parton distribution \cite{FPPS}. 
  Indeed, the product-form model
 has a $k_{\perp}^4$  fall-off of the BSA, compatible
 with a BSA kernel dominated by the  one-gluon-exchange (OGE) and
 faster than the ${k}_{\perp}^2$ decay of the sum-form model,
  and both at $~-t$ $\rightarrow \infty~$ and at $~x \rightarrow 0~$ or 
 at $~x \rightarrow 1~$ the high
 momentum part of the pion state is probed.
 
 



At $~|\xi|=1~$ and at $~\xi=0~$ the
covariant product-form model exhibits the same general behavior of the other models and 
for all of the considered models at high $t$ 
a maximum of GPD's around $x \sim 1$ occurs \cite{FPPS}. 
This last fact has a
simple kinematical explanation at $~\xi=0~$.
At $~\xi=-1~$, where only the $q \bar{q}$ pair production contributes,
one has $x = 2k^+/\Delta^+$, while
 large
$|t|$ values mean large $\Delta^+ = \Delta_z
 \approx 2 k_{zq}$, and $2 k^+ = k^+_q - k^+_{\bar{q}} 
 \sim 2 E_q = 2 ~ \sqrt{m^2 + {\bf{k}}^2}$, since each quark in the pair is almost on its mass
 shell. Then
 $ x \sim E_q / k_{zq} ~~ \rightarrow  ~~ 1 ~ .$
As noticed in \cite{FPPS}, also at  $|\xi|=x$ 
 the maximum of GPD as $-t ~ \rightarrow ~ \infty$ moves from $~x \sim 0.5~$ towards
$~x = 1$.  The GPD
at $|\xi|=x$ allows one to
 explore the transition from the valence to the NV region and this kinematical
regime should be relevant to study single spin asymmetry \cite{diehlpr}.

\vspace{-0.cm}
\section{Nucleon 
parton momentum distributions}
\label{PaceE_sec:10}
We describe the quark-nucleon vertex function through a BSA,
with a Dirac structure suggested
by an effective Lagrangian \cite{de},
and adopt a Breit reference frame where~$~~{\bf q}_{\perp}=0$ and
$q^+=
|q^2|^{1/2}$. Our CQ mass is ~$m=m_u = m_d = 200 ~ {\rm{MeV}}$.
 The  
 current in the SL region
 is approximated {\em{microscopically}}  by the Mandelstam formula \cite{mandel}
\vspace{0.0cm}
\be
\hspace{-0.9cm}\langle \sigma',P_N'|j^\mu~|P_N,\sigma\rangle
=3~N_c  
 \int {d^4k_1 \over (2\pi)^4}\int {d^4k_2 \over (2\pi)^4}  
\sum \left \{ \bar \Phi^{\sigma'}_N(k_1,k_2,k'_3,P_N')
 S^{-1}(k_1)  S^{-1}(k_2)~{\cal I}^\mu(k_3,q)~
 \Phi^\sigma_N(k_1,k_2,k_3,P_N)\right \}
\ee
where 
$k'_{3} = k_{3} + q$, ${\cal I}^\mu(k_3,q)$
is the quark-photon vertex, and $\sum$ implies a sum over isospin and
spinor indexes. 

The 
Mandelstam formula is projected out by an analytic integration on $k_1^-$ and  $k_2^-$,
 taking into account only the poles of the propagators. 
Then the current becomes the sum of a purely valence contribution 
and a NV, pair-production contribution.
Clearly, after the $k^-$ integrations,
the vertex functions depend only upon the LF three-momenta.

The quark-photon vertex has IS and IV contributions,
$ {\cal I}^\mu=~{\cal I}^\mu_{IS} +\tau_z
  {\cal I}^\mu_{IV}
  \label{curr}
 $,
and each term  contains a purely valence contribution 
(in the SL region only) and a contribution corresponding to the pair production (Z-diagram). 
  In turn the Z-diagram contribution can be decomposed in a bare term $+$  a
  VMD term, viz
\be
\hspace{-0.7 cm} {\cal I}^\mu_{i}(k,q) = {\cal N}_{i} ~ \theta(P_N^+-k^+) ~ \theta(k^+) ~
   \gamma^\mu + \theta({q}^+ + k^+)
   ~
 \theta(-k^+)~\left \{ Z_B~{\cal N}_{i} ~ \gamma^\mu+ 
  Z^i_{VM}~\Gamma^\mu(k,q,i)\right\} 
   \ee
 with  $i = IS, IV$, ${\cal N}_{IS}=1/6$ and ${\cal N}_{IV}=1/2$.   The constants $Z_B$ 
 and 
 $Z^i_{VM}$ 
 are unknown weights to be extracted from
 the phenomenological analysis of the experimental data.
 According to the label $i$, the VMD term $\Gamma^\mu(k,q,i)$, 
 which does not involve free parameters, 
  includes IV or IS mesons. Indeed in \cite{nucleon}
   the microscopic model for the VMD, successfully used in
    \cite{DFPS} for the pion ff 
  and based on the  mass operator of Ref. \cite{Fred},
   was extended to IS mesons.
 \begin{figure}[t]
\vspace{-1.5cm}
\hspace{-.0cm}{\includegraphics[width=6.5cm]{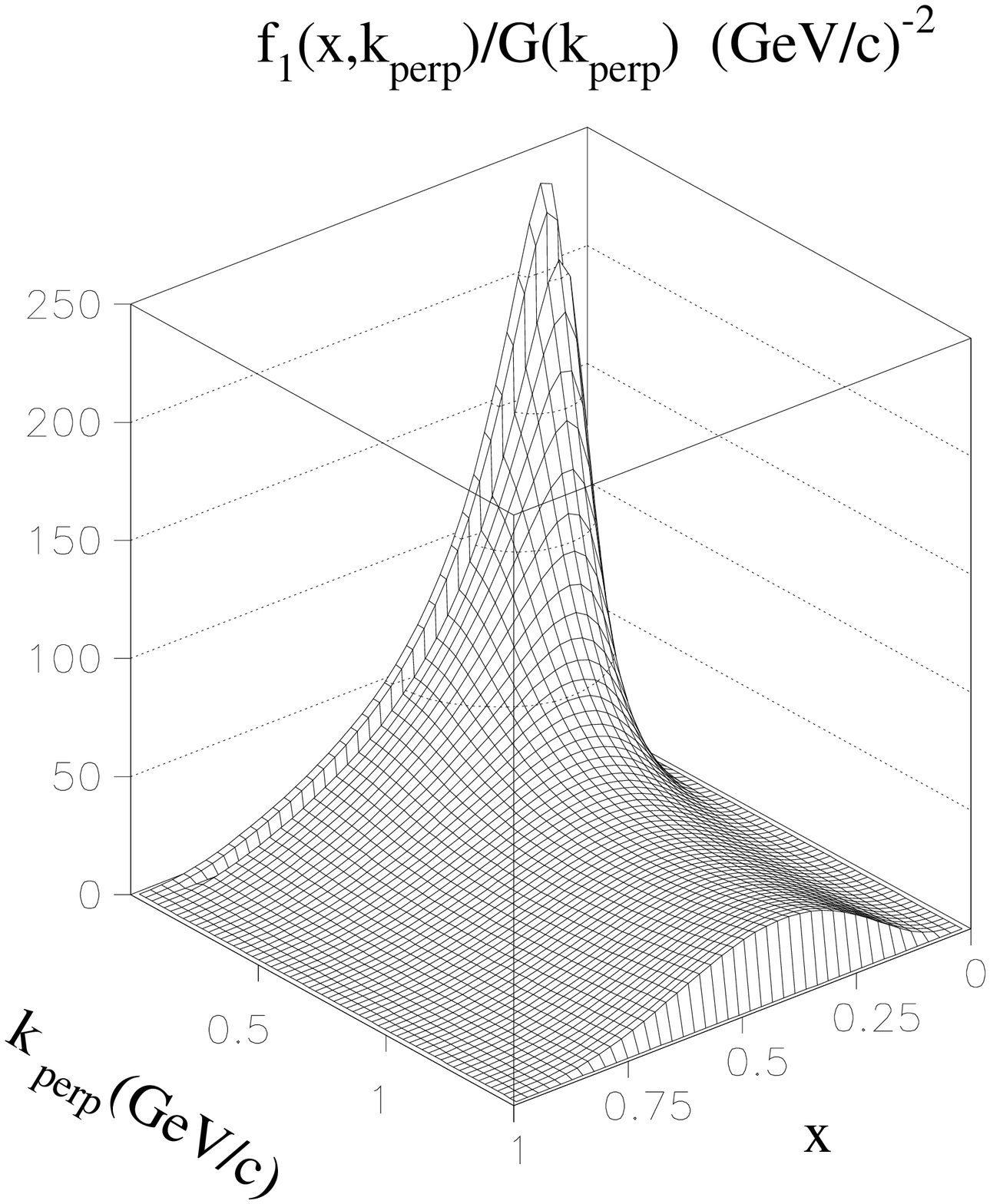}}
\hspace{.7cm}
{\includegraphics[width=6.5cm]{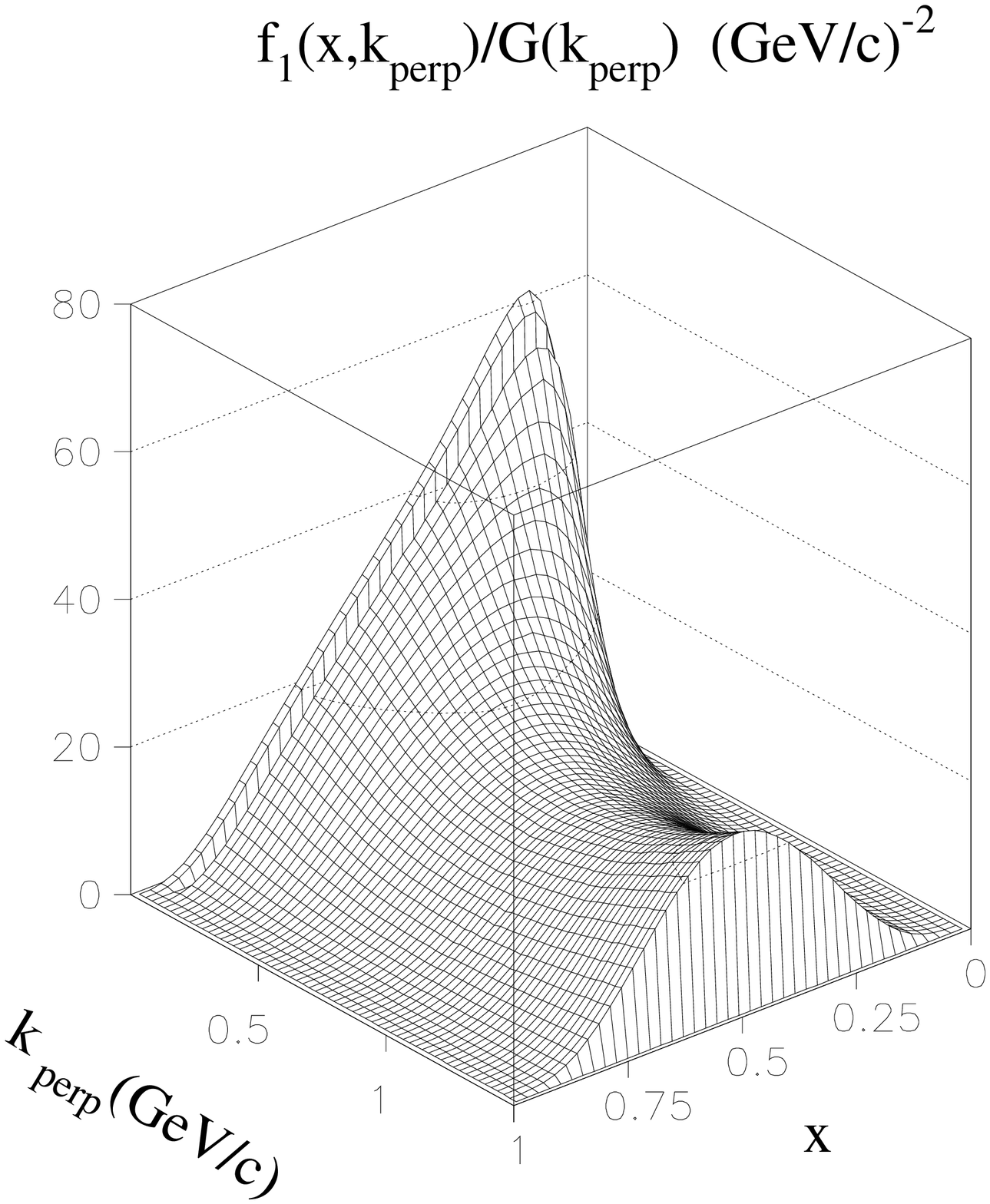}}
\vspace{-1.7cm}
\caption{Left panel: transverse-momentum distributions for a $u$ quark in  
the proton. $G(k_{\perp}) = (1 ~ + ~ k_{\perp}^2/m_{\rho}^2)^{-5.5}$,
$m_{\rho}$ = 770 MeV and $k_{perp} =  k_{\perp} $. Right panel: the same as in 
the left panel, but for a $d$ quark inside  
the proton (after Ref. \cite{PDFPS}). } 
\end{figure} 

  In the valence vertexes  the spectator quarks are on the
$k^-$-shell, and the BSA momentum dependence  
is approximated through a nucleon wf PQCD inspired, 
which depend on 
 the free mass of the three-quark system,  $M_0(1,2,3)$,
\be
\hspace{-0.6cm} \Psi_N(\tilde{k}_1,\tilde{k}_2,P_N)  = P_N^+ ~
{ {\Lambda}(k_1,k_2,k_3)|_{(k_{1on}^-,k_{2on}^-)} \over [m_N^2 - M^2_0(1,2,3)]^{~}}  
 = ~  P_N^+ ~ {\cal {N}}~
{~(9~m^2)^{7/2}
 \over (\xi_1\xi_2\xi_3)^{p}~ \left[\beta^2 + M^2_0(1,2,3)\right]^{7/2}} \quad ,
\ee
where $\tilde{k}_i \equiv (k_i^+,{\bf k}_{i\perp})$, 
$\xi_i = k^+_i/P_N^+$ ~($i=1,2,3$)
and  ${\cal N}$ is a normalization constant.
The power $ 7/2 $ and the parameter $ p = 0.13 $ are chosen to have
an asymptotic decrease of the triangle contribution faster than the dipole.
Only the triangle diagram determines
the magnetic moments, which are weakly dependent on $p$. Then
$\beta = 0.645$ can be fixed by the magnetic moments and we obtain $\mu_p = 2.87\pm0.02$ ~
($\mu_p^{exp}$ = 2.793)
and $\mu_n = -1.85\pm0.02$ ~ ($\mu_n^{exp}$ = -1.913).

For the Z-diagram contribution, the NV vertex 
is needed.
 It can depend on  the free mass, $M_0(1,2)$, of the (1,2) quark pair 
and on the free mass, $M_0(N,\bar {3})$, of the ( nucleon - quark $\bar {3}$ ) system 
entering the NV vertex.
Then in the SL region we approximate the
 momentum dependence of the NV vertex 
 $ {\Lambda}_{NV}^{SL} = {\Lambda}(k_1,k_2,k_3)|_{k^-_{1on},k^{'-}_{3on}}$ by
${\Lambda}_{NV}^{SL}= [g_{12}]^{2}[g_{N\bar {3}}]^{3/2}
 \left [{k_{12}^+ / P_N^{\prime +} }  \right ]
  \left [ P_N^{\prime +} / k_{\overline {3}}^+ \right ]^r
  \left [P_N^{+} /  k_{\overline {3}}^+   \right ]^{r} 
  $, 
with
$
 k_{12}^+ = k_1^+ +  k_2^+ $ and $  g_{AB} = (m_A ~ m_B) /\left
[\beta^2+M^2_0(A,B)\right] $.
The power 2 of $[g_{12}]^{2}$ is suggested from counting rules. 
The power 3/2 of $[g_{N\bar {3}}]^{3/2}$ and the parameter $r=0.17$ are chosen
 to have an asymptotic dipole behavior for the NV contribution,
 as suggested by the OGE dominance.

 We performed a fit for the ff's of our free parameters, 
 $Z_B$, $Z^i_{VM}$, $p$, $r$  in the SL region, obtaining
 a $\chi ^2$/datum = 1.7. 
The Z-diagram turns out to be essential 
in our reference frame
with $q^+ > 0$. 
In particular, {\em{the possible zero in $G_E^p/G_M^p$ for $q^2<0$ is strongly related 
to the pair-production contribution, i.e. to higher Fock state components}}.


The longitudinal distribution $q(x)$ is the limit in
     the forward case  of the unpolarized 
     GPD ${H}^q(x,\xi,t)$.
For $P_N' = P_N$, both $q^+$ and $\xi$ are vanishing and $x = k_3^+ / P_N^+ = \xi_3$
is the fraction of the active quark longitudinal momentum.
 As a consequence the function ${H}^q(x,\xi,t)$ 
reduces to the longitudinal parton
distribution function $q(x)$:
\begin{eqnarray}
\hspace{-1cm}  {H}^q(x,0,0) = q(x) = \int d{\bf k}_{\perp} ~~ f_1^q(x,k_{\perp})  ~
 = ~ \int \frac{dz^-}{4\pi} e^{i x P^+ z^-} 
 \left . \langle P_N | \bar \psi_q
(-\frac{z}{2}) \gamma ^+ \,  \psi_q (\frac{z}{2}) | P_N \rangle \right |_{\tilde{z}=0} ~~ ,
  \label{struc1} 
  \end{eqnarray}
    where 
    an average on the nucleon helicities is understood.
   Once all the parameters of the nucleon light-front wf 
   $\Psi_N(\tilde k_1,\tilde k_2,P_N)$ have been determined, one can easily define
 the TMD of the active quark, $f_1^q(x,k_{\perp})$,  in terms of the  LF wf
and through Eq. (\ref{struc1})
 also  the longitudinal  distribution of the struck quark. From the isospin symmetry 
 one has
$~u_p(x)=d_n(x)=u(x)$ and
$d_p(x)=u_n(x)= d(x)$.

Our preliminary results for $f_1^{u(d)}(x,k_{\perp})$ in the proton and for $u(x)$ 
and $d(x)$  are
shown in Fig. 3  and in Fig. 4, respectively.
It can be observed that the decay of our $f_1$ vs $k_{\perp}$ 
is faster than in diquark models of the nucleon
\cite{Jacob}, while it is slower than in gaussian factorization
models for the TMD \cite{Anselm}.
As far as the longitudinal  distributions are concerned, 
a reasonable agreement of our $u(x)$  with the CTEQ4 fit to the experimental data 
\cite{Lai} 
can be seen in Fig. 4.

\section{Conclusions}
\label{PaceE_con}
 Microscopical models for pion 
and  nucleon em form factors  
 have been investigated 
  with  good results in the SL region. 
  The Z-diagram (i.e. higher Fock state component) has been shown to be essential, 
in   reference
frames  where $q^+ \ne 0$.

 The analysis of the form factors allows us to get hadron vertexes that are used to
 evaluate the unpolarized longitudinal  and transverse momentum  distributions.
 The effects of a $k_{\perp}$ fall-off  
 compatible with the OGE dominance has been explored. Its  role 
 for the description  of the ff tail  at high $-t$ and of vanishing 
  longitudinal parton distributions at the end
 points has been shown.

The covariant product-form model is able to reproduce the pion GPD's evaluated 
  by the
Mandelstam-inspired model at $|\xi|=1$ and by the LFHD model at $\xi=0$. Then
 one could argue that the product-form
 model contains the main 
ingredients for the description of the constituents inside the pion and could be applied 
to study
experimental data.

 Our next step will be the calculation of polarized  longitudinal and transverse 
 momentum distributions.

 \vspace{.2cm}
\begin{figure}[t]
\centering
\vspace{.8cm}
\hspace{.0cm}{\includegraphics[width=6.5cm]{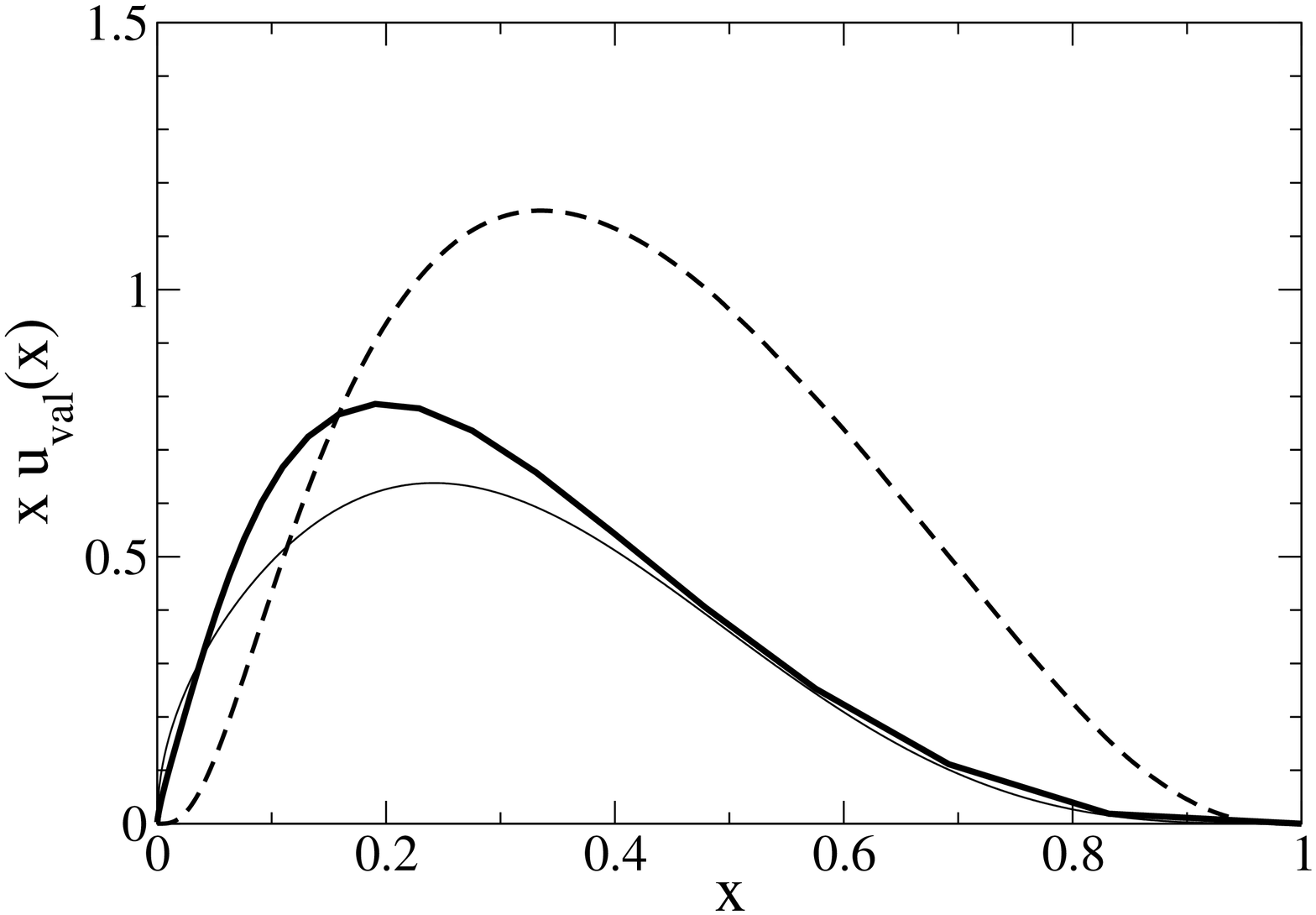}}
\hspace{.7cm}{\includegraphics[width=6.5cm]{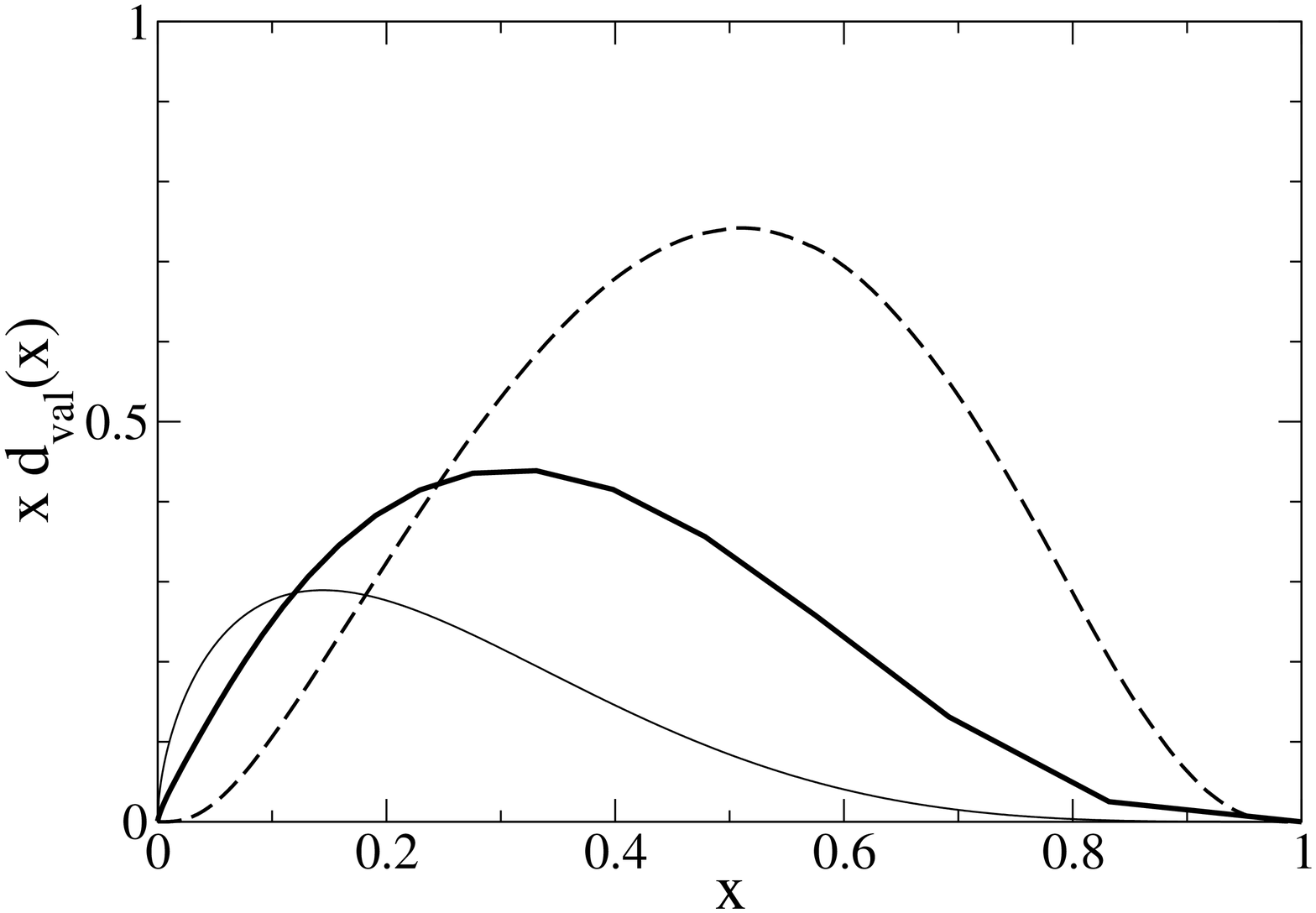}}
\vspace{.0cm}
\caption{Left panel: longitudinal momentum distribution for a $u$ quark inside  
the proton. Dashed lines: our preliminary results; thick solid lines: our results after
evolution to $Q^2$ = 1.6 (GeV/c)$^2$; 
thin solid lines: CTEQ4 fit to the experimental data \cite{Lai}. Right panel: 
the same as in the left panel, but for a $d$ quark in 
the proton (after Ref. \cite{PDFPS}). } 
\end{figure}





\bibliographystyle{aipproc}   





\end{document}